\input epsf.tex

\documentstyle[prl,aps,twocolumn,floats]{revtex}
\begin{document}
\title{Basic considerations on broken-symmetry states of organic superconductors
 }
\author{C. Bourbonnais}
\address{Centre de Recherche sur les Propri\'et\'es \'Electroniques de
Mat\'eriaux Avanc\'es, D\'epartement de Physique,}
\address{Universit\'e de Sherbrooke, Sherbrooke, Qu\'ebec, Canada J1K 2R1}

\maketitle
\begin{abstract}
In this brief account, we shall review   the  different mechanisms  giving rise to  itinerant
and localized 
  antiferromagnetism   in  quasi-one-dimensional organic conductors: the
Bechgaard salts and their sulfur analogs. We will then focus on the problem  of spin correlations and their
impact on the pairing mechanism  for organic superconductivity.  
\end{abstract}

\section{Introduction}
The discovery of superconductivity   twenty years ago by J\'erome {\it et al.}\cite{Jerome80}, in the
charge transfer salt  (TMTSF)$_2$PF$_6$  was not only remarkable for superconductivity 
itself, but  also  showed that superconductivity in a quasi-one-dimensional electronic
material could emerge  from a
competition with antiferromagnetism under pressure. This unusual interplay between magnetism and
superconductivity  
 turned out  to be a  commom characteristic of other members of   the Bechgaard salts series (TMTSF)$_2$X
and their sulfur analogs: the Fabre salts (TMTTF)$_2$X \cite{Jerome82}.  Similar competition of ground states
was also found in other low-dimensional many-electron systems such as the layered  organic superconductors and 
high-T$_c$ cuprates \cite{Kanoda97,McKenzie98,Lefebvre00}.

 Antiferromagnetism and superconductivity  are therefore closely related in these systems and can no longer
be considered as completely blind to each other. In spite of the considerable interest  bestowed on  this issue
over the last twenty years, their connection remains so far largely unexplained and stands as  one of the most
important challenges of the physics of  low-dimensional organic superconductors.   

\section{Antiferromagnetism}

\begin{figure}[htb]
\epsfxsize 8 cm
\centerline{\epsfbox{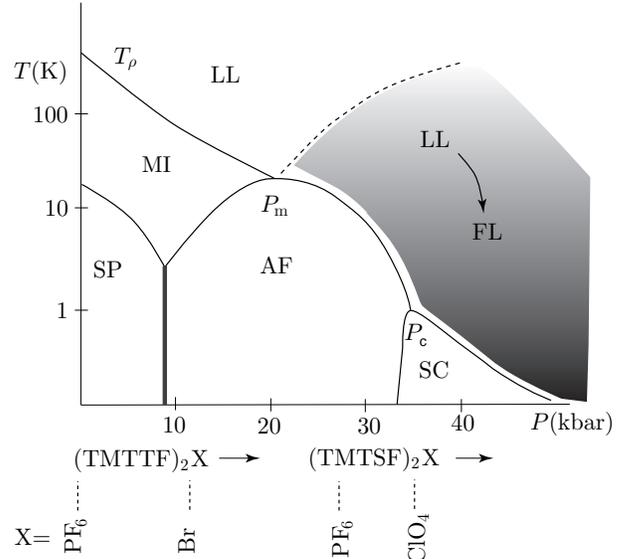}}
\caption{The generic  phase diagram of the Bechgaard and Fabre salts [(TM)$_2$X] as a function of pressure or
anion X substitution, after ref.\protect\cite{Bourbon99}.  }
\label{Diagramme}
\end{figure}

By sharing  a common boundary  with superconductivity 
and being  present in a relatively large domain of the normal phase in the form of short range correlations,
antiferromagnetism has become a dominant feature of the whole phase diagram of (TM)$_2$X 
(Figure~\ref{Diagramme}). Thus attempts  to unravel  the origin of superconductivity  in the
Bechgaard salts  and their sulfur analogs  must inevitably go through a proper  understanding  of  the mechanisms
leading to magnetism in these  electronic materials. 

\subsection{The nesting mechanism: success and limitations}
 The observation of itinerant antiferromagnetism  in (TMTSF)$_2$X  indicated from the outset that
electron-electron Coulomb repulsion,  not electron-phonon interaction, would play a more   important part in
the instability of the metallic state.  However, it should be stressed that repulsive interactions alone, only
lead in a
metal to uniform, (i.e., ferromagnetic) spin correlations. These  enhance  the magnetic spin susceptibility 
with respect to the Pauli or noninteracting band limit $-$ an effect that is confirmed by experiments
\cite{Bourbon99}. Adopting a weak coupling picture for electrons, the mechanism of density-wave formation, called
nesting, originates  from an instability of the    Fermi surface. For a quasi-one-dimensional  metal, 
 constant energy electron and hole states  located   on opposite sides ($\sim\!\pm {\bf k}_F$)   of the
open Fermi surface can be connected or nested on one another. This is expressed  by the following  relation for
the energy of one-electron states
\begin{equation}
\epsilon_\pm({\bf k}) = - \epsilon_{\mp} ({\bf k} + {\bf Q}_0) + \delta\epsilon_0,
\end{equation}
 where  ${\bf Q}_0=(2k_F^0,{\bf q}^0_\perp)$ is
called the nesting vector of the Fermi surface,  with $2k_F^0$ corresponding to the nesting vector of the 
single chain. Here, $\delta\epsilon_0$ stands as   corrections that  parametrize  deviations from perfect nesting.

For  perfect nesting, the electron gas develops a singular logarithmic response to  density-wave
formation  at wavevector ${\bf Q}_0$, as shown by the expression of the free electron gas susceptibility at ${\bf
Q}_0$:
\begin{eqnarray}
\chi_0({\bf Q}_0,\omega) =  && {2\over LN_\perp} \sum_{{\bf k}} {n[\epsilon_+({\bf k})] -n[\epsilon_-({\bf k}+{\bf
Q}_0)]\over \epsilon_+({\bf k}) - \epsilon_-({\bf k}+{\bf Q}_0) -\omega + i0^+} \cr
                      = && N(E_F) \ln {E_0\over T} \ \ \ \ \ (\delta\epsilon_0=0,\omega=0),
\label{chi}
\end{eqnarray}
where the last expression follows in the satic limit. Here, $E_0$ is a cut-off energy and $N(E_F)$ is the
noninteracting electron density of states at the Fermi level. Repulsive   interaction 
that scatters
 two electrons at $\pm {\bf k}_F$  gives rise to an attraction between
  electron and hole  separated by ${\bf Q}_0 $,   which according to (\ref{chi}) will be  
 magnified as function of temperature and can become singular. In leading (ladder) order, this is well
known  to yield the simple pole singularity of the total forward scattering amplitude, namely
\begin{equation}
\Gamma({\bf Q}_0,\omega) = {g^* \over 1- g^* \chi_0({\bf Q}_0,\omega)},
\label{ladder}
\end{equation}
where $g^*$ is an effective  coupling defined at the scale
$E_0$ (for simplicity, we have dropped the contribution of umklapp
scattering). The characteristic temperature
$T_{SDW}$ signaling an SDW instability of the normal state  then follows  the pattern  of the BCS mechanism for
superconductivity.
 The temperature at which  bound electron-hole pairs condense
 is determined by the condition
\hbox{$1=g^*\chi_0({\bf Q}_0,T_{SDW})$}, corresponding to  the simple pole singularity of $\Gamma$ in the static
limit, and this yields
\begin{equation}
T_{SDW} \sim t_\perp^* e^{-1/N(E_F)g^*}.
\end{equation}
Within this weak-coupling  picture of the SDW transition, the identification of the cut-off
$E_0$ in (\ref{chi}) with an effective $-$ renormalized $-$ value of the interchain hopping  amplitude $t_\perp^*
(< t_\perp)$ takes on particular significance  in quasi-one-dimensional systems. It stands as the
temperature for crossover towards  the one-dimensional metallic state.   
It imposes a mechanical limitation of the above approach to the temperature domain  $T<t_\perp^*$, where the
curvature of the Fermi surface and in turn the transverse momentum of electrons  is coherent in the
quantum-mechanical sense. Otherwise, for
$T> t_\perp^*$,  thermal fluctuations reduce  nesting properties to be  essentially longitudinal or
one-dimensional in character, a configuration  that introduces interference between  electron-hole  and
electron-electron (Cooper) pairings. An immediate outcome of this interference is to invalidate the ladder
summation in (\ref{ladder}), which assumes that electron-hole pairing can be singled out in  perturbation
theory. The metallic state then turns out to be no longer a Fermi liquid but  rather a Luttinger liquid.  For
repulsive interactions, however,  SDW correlations are singular over the whole 1D temperature domain. According
to the one-dimensional theory,
  their amplitude  is governed by a  power law
increase of  the susceptibility  at wavevector $2k_F^0$:
\begin{equation}
\chi_0(2k_F^0, T) \sim T^{-\gamma}, 
\label{power}
\end{equation}
 where $\gamma$ is an exponent that depends on the strength of electron-electron
interaction. 
 
The application of pressure modifies electron band energy  in such a way that nesting  deviations ($\delta
\epsilon_0$)  increase and cut off  the logarithmic singularity following the approximate expression 
\begin{equation}
\chi({\bf Q}_0,T) \approx  N(E_F) \ln {t_\perp^*\over \sqrt{\delta\epsilon_0^2 + T^2} }.
\end{equation}
As pressure increases, the SDW critical temperature derived from (\ref{ladder})  is then progressively shifted
down to lower temperature  and ultimately vanishes  whenever  $\delta\epsilon_0(P)$ exceeds a certain threshold
($\sim T_{SDW}$) \cite{Yamaji82}. The rapid suppression of $T_{SDW}$ thus obtained under pressure  
 squares relatively well with  observation made close to $P_{c2}$  (Figure
\ref{Diagramme}). The involvement of  nesting as a mechanism of  stability of itinerant antiferromagnetism is
however most substantiated  by its remarkable ability to account for the restoration of a cascade of SDW
states induced  by a magnetic field 
above $P_{c2}$ \cite{Gorkov84,Heritier84}. When a magnetic field is applied along the less conducting axis, open
orbits are produced which confine  electron motion preferentially along the chains where full, but quantized,
nesting properties are found to  be restored. In this way, the logarithmic singularity in $\chi_0({\bf Q}_N,T,H)$
is recovered  for a given set of discrete nesting  wavevectors
${\bf Q}_N$  and shows a particular variation  with field. 
Following (\ref{ladder}), a characteristic hierarchy or cascade of  $T_{SDW}(N)$ is found with quite remarkable
consequences on physical properties. For example, the quantization
of Hall resistance has received sound experimental support \cite{Jerome94,Chaikin96} .

\begin{figure}[htb]
\epsfxsize 8 cm
\centerline{\epsfbox{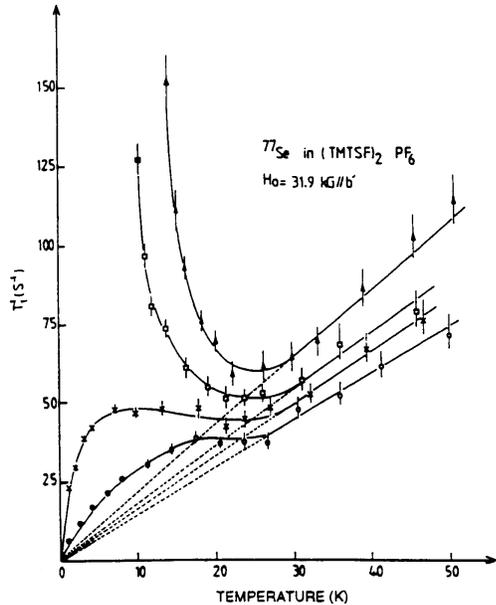}}
\caption{$^{77}$Se NMR relaxation rate as a function  of temperature below (1bar and 5 kbar) and above (8 kbar
and 10 kbar) $P_c$ in (TMTSF)$_2$PF$_6$, after ref. \protect\cite{Creuzet87b}.}
\label{Relaxation}
\end{figure}

In spite of this apparent success of the  BCS mechanism to explain long-range SDW order, difficulties arise when 
one tries to tackle from the same  standpoint the properties of the metallic state. One of the main defects of
the approach, as it is formulated,  lies in the assumption that all electron excitations of the normal state can
be described  within the  Fermi liquid framework,  which is not corroborated by most experiments. In a Fermi
liquid for example, electron correlation effects are essentially absent of the scene from  $T\sim t_\perp^*$ down
to  the close vicinity of
$T_{SDW}$, where the onset of critical fluctuations ultimately limits the validity of the quasi-particle picture.
Thus in the BCS description, the  temperature interval $\Delta T_{fl}$ for SDW fluctuation effects, albeit
increased by  spatial anisotropy,  is small compared to $T_{SDW}$ and a  reduction of $T_{SDW}$ is
concomitant to a reduction of $\Delta T_{fl}$, which vanishes with $T_{SDW}$. 

The  NMR spin-lattice relaxation rate
$T_1^{-1}$ was among the first properties to  show the existence in the Bechgaard salts of spin correlations 
 whose amplitude did not agree with the Fermi liquid prediction
\cite{Creuzet87b,Bourbon84}. In this matter, the  temperature dependent
$T_1^{-1}$ data of Figure~\ref{Relaxation} obtained for (TMTSF)$_2$PF$_6$ are particularly revealing of the
importance of antiferromagnetic spin correlations in the normal state  near
$P_{c}$. For  pressures above $P_{c}$, where  the system is no longer in a SDW state but is expected to be a
superconductor,  a huge enhancement of the relaxation rate  persists over a  large temperature
interval. Given the extent in temperature  of the enhancement and its non critical 
character, it contrasts  with the  Korringa law behavior $(T_1T)^{-1} \approx
$~constant, that should be found for a weakly interacting Fermi liquid.  In view of quite similar effects reported
for other members of the series, as well as other experimental signs of correlation effects \cite{Bourbon99}, all
this goes to show that antiferromagnetic
spin correlations of the normal state of (TMTSF)$_2$X differ from those
found in a  conventional metal. 

On the theoretical side, it is worthy of note that it was at first proposed that these non Fermi
liquid effects are attributable to a sizeable reduction
$-$ or renormalization $-$ of 
the one-dimensional temperature scale 
$t_\perp^*$ as a self-consistent effect of correlations themselves \cite{Bourbon84}. According to this scheme, a
reduction of $t_\perp^*$  would  widen the
 temperature domain  in which low-dimensional spin correlations  lead to  a 
power law  increase of  enhancement for  the relaxation rate $(T_1T)^{-1} \sim T^{-\gamma}$. This
  renormalization effect on the kinetics of electrons   does solve a number of problems about
the normal state but is not beyond creating new paradoxes \cite{Georges00} $-$ in particular concerning the 
 transport properties  and field-induced SDW states \cite{Danner94,Moser98,Zheleznyak99}. 
  A central  point at issue is the determination of the actual value taken by this temperature scale  and
 if $t_\perp^*$   rather corresponds to a smooth crossover between 
the Luttinger  and Fermi liquids (Figure~\ref{Diagramme}).

\subsection{The sulfur series and the Mott insulating picture}
It is instructive  to compare the Bechgaard salts with the (TMTTF)$_2$X series of Fabre
salts. The members of this  series are located  on the left of  the  phase
diagram in  Figure~\ref{Diagramme}; their study has provided  extremely valuable and rather well
understood pieces of information about both the nature of correlations and the mechanisms of long-range order
 in a sector of the phase diagram that stands  only few kilobars apart from the Bechgaard salts. At low
pressure, the nature of the transition in (TMTTF)$_2$X  shows qualitative differences with respect to the
Bechgaard salts: antiferromagnetism    is stabilized from a paramagnetic  state that is  insulating. The
metal to insulator `transition' occurs at a characteristic  temperature usually denoted as
\hbox{$T_\rho
\sim 10^2$~K }(Figure~\ref{Diagramme}), below which electrical transport develops a thermally activated behavior
\cite{Coulon82}. Since  spin degrees of freedom remain unaffected at $T_\rho$ $-$ as shown by the absence of
anomaly in the spin susceptibility at
that temperature
\cite{Bourbon99} $-$ (TMTTF)$_2$X are good examples  of Mott insulators below
$T_\rho$ \cite{Emery82}. The measured amplitude of the Mott insulating gap
\hbox{$\Delta_\rho\sim 3T_\rho$} even exceeds the transverse  bare hopping
amplitude
$t_\perp$ obtained from band calculations \cite{Canadell94}. The  carriers can  then  be considered as  confined
on organic stacks, while spin excitations  are gapless  and dominated by one-dimensional physics.  The
one-dimensional nature of spin correlations  associated to the Mott insulating state in (TMTTF)$_2$X has
received  several experimental confirmations  
\cite{Creuzet87,Wzietek93}.   As regards to NMR for example, the one-dimensional theory predicts 
a  power law enhancement  of
\hbox{$(T_1T)^{-1} \sim T^{-\gamma}$} at
$2k_F^0$ with the exponent \hbox{$\gamma=1$} for a 1D Mott system  \cite{Bourbon93,Emery79}, a behavior
that has indeed been found
 in the paramagnetic phase of all sulfur compounds studied down to the vicinity of their
critical point  
\cite{Wzietek93}.

\begin{figure}[htb]
\epsfxsize 8 cm
\centerline{\epsfbox{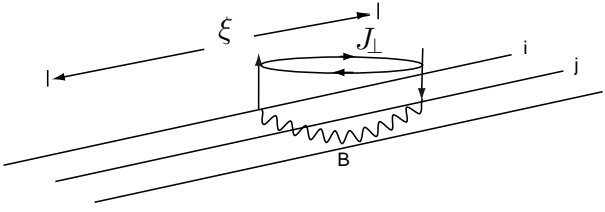}}
\caption{Interchain  kinetic exchange (superexchange) between spins on  neighboring
chains
$i$ and
$j$. In the presence of a Mott gap, the antiferromagnetic exchange takes place over the coherence length
$\xi\to\xi_\rho
\sim v_F/\Delta_\rho$ corresponding to the size of bound electron-hole pairs along the chains. The process given
in B refers to exchange of spin fluctuations which gives rise to attractive pairing on different chains.    }
\label{exchange}
\end{figure}

 In spite of charge confinement along the chains,
spin degrees of freedom ultimately deconfine and lead  to antiferromagnetic long-range
order which is found in the range  5-25~K  for (TMTTF)$_2$X. However, the mechanism by which
antiferromagnetic correlations propagate in the transverse direction is here not connected with the Fermi surface.
Owing to charge localization, the system is clearly not a Fermi liquid so that the antiferromagnetic transition
cannot be  drived from  Fermi surface effects but rather from the interchain kinetic exchange (superexchange) 
coupling
$J_\perp$ \cite{Brazovskii85b,Bourbon86}. This  coupling between spins
  is in effect not present in the Hamiltonian in the high-energy limit; it  is
generated
  through virtual interchain hopping of electrons  confined within the coherence length \hbox{$\xi_\rho \sim
v_F/\Delta_\rho$} induced by the Mott  gap (Figure~\ref{exchange}). The explicit expression for
the effective interchain exchange Hamiltonian at the scale \hbox{$\sim T_\rho$} is known and can  be determined by
the renomalization group method
\cite{Bourbonnais88,Bourbon95}: 
\begin{equation}
H_\perp \approx \int dx \sum_{\langle i,j\rangle} J_\perp {\bf S}_i(x)\cdot {\bf S}_j(x),
\label{Jperp}
\end{equation}
where
\begin{equation}
J_\perp \approx {\xi_\rho\over a} {t_\perp^{*2}\over \Delta_\rho}.
\label{Jperp_ampl}
\end{equation}
Here $t_\perp^*$ stands as the effective interchain hopping taking place at the energy scale $\sim T_\rho$.
$J_\perp$ couples to singular antiferromagnetic correlations along the chains and induces a  true transition
at finite temperature (in three dimensions). In a very good approximation, the transverse part $H_\perp$ can be
treated in molecular-field approximation and 1D correlations exactly, so that the singular part of the total
antiferromagnetic susceptibility at the
$(2k_F^0,\pi)$ reads
\begin{equation}
\chi({\bf Q}_0,T) = {\chi(2k_F^0,T) \over 1 - J_\perp \chi(2k_F^0,T)},
\end{equation}
where $\chi(2k_F^0,T)$ is given by the one-dimensional form (\ref{power}) with $\gamma=1$. The N\'eel transition
then takes place 
at
\begin{equation}
T_N\sim  {t_\perp^{*2}\over \Delta_\rho}.
\end{equation}
It follows  from this result a characteristic increase of $T_N$ as 
$\Delta_\rho$ or
$T_\rho$ decreases under pressure. This  feature which comes from a magnification of
$J_\perp$  due to the  progressive deconfinement of carriers,  agrees with observations  (cf. Figure
\ref{Diagramme})
\cite{Klemme95}. In the case of (TMTTF)$_2$Br for example, a  relation of
the form $T_\rho T_N \approx$~constant under pressure has been shown to be well satisfied by the data
\cite{Brown97}. 

The drop of $T_\rho$ carries on under pressure until it merges with the critical domain associated with the
transition. This pressure scale, which is denoted as $P_m$ in  Figure~\ref{Diagramme}, 
signals a change of regime of the normal state, which becomes metallic. The insulator to metal crossover  marks
the onset of charge deconfinement and weaker coupling conditions for the carriers.  This is exemplified by a
reduction of the enhancement in NMR $(T_1T)^{-1}$ \cite{Wzietek93},  and  the restoration  of a transverse plasma
edge in optical experiments\cite{Vescoli98}. Such a change in the strength of electron correlations has a
characteristic impact on the pressure profile of $T_N$ under pressure, which presents a maximum at $P_m$. This
feature, which is predicted by microscopic calculations \cite{Bourbon95}, comes essentially from the pressure
variation of
$J_\perp$ whose magnification in  (\ref{Jperp_ampl}), as a result of the increase of  $\xi_\rho $,
levels off at the boundary and finally decreases above
$P_m$. It turns out that even in the absence of an insulating behavior, $J_\perp$ remains the main driving
mechanism of the transition close to
$P_m $. Its influence grows  less and less as one moves away from $P_m$ to finally become coupled to
the nesting mechanism whose importance follows the growth of the Fermi liquid component under pressure. As one
can see, the question that was previously raised  about both the degree of renormalization of $t_\perp^*$ and
correlation effects in the normal phase of the Bechgaard salts, is closely related to  the   extent to which
antiferromagnetic exchange is still active for magnetic ordering on  the  right-hand side of the diagram of
Figure~\ref{Diagramme}.  

\begin{figure}[htb]
\epsfxsize=0.9\hsize
\centerline{\epsfbox{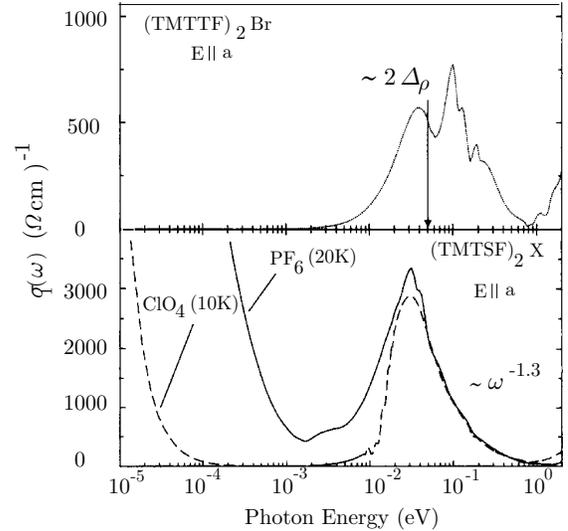}}
\caption{One-chain optical conductivity of members of (TM)$_2$X in the normal phase at low temperature. The
arrow indicate the size of  (twice) the Mott gap, after \protect\cite{Vescoli98}.  }
\label{Conduct}
\end{figure}

On the theoretical side, this important question has  not received a clear answer so far. Nonetheless, 
it is instructive in this matter to look at experiments giving  relevant information about the spectral response
of these materials\cite{Vescoli98,Timusk96}.   For example, the optical conductivity obtained by Vescoli {\it et
al.}\cite{Vescoli98} from infrared reflectivity data,  are shown in Figure~\ref{Conduct}  for members  of both
series in their normal phase.  
The results show that the Mott gap, which is the dominant structure in the sulfur compound,  remains  clearly
visible in the Bechgaard salts, yet metallic
\cite{Giamarchi97}. The gap structure  captures 
  99\% of the spectral weight,  whereas the large metallic DC conductivity  provided by a narrow frequency mode
carries the remaining 1\%. These results clearly indicate that  the build-up of short-range correlations
along the stacks is achieved for strongly coupled  electrons. This implies in turn that interchain exchange is
likely to be well developed in (TMTSF)$_2$X. 

\section{Organic superconductivity}  
We now examine the impact  spin correlations can have on the nature of organic
superconductivity. We ground the  analysis on the following experimental information: i) as shown by NMR, the
normal phase is dominated by antiferromagnetic correlations (Figure~\ref{Relaxation}); ii) the superconducting
$T_c$ and
$T_{SDW}$ join at
$P_c$, where
$T_c$ presents a maximum (Figure~\ref{Diagramme}); iii) there is no  lattice softening at $2{\bf k}_F$. What
should be established at the very outset is that spin correlations in the normal phase  lead to an
enhancement of electron-electron  repulsion.  This immediately follows  from the expression of
the vertex part in (\ref{ladder}) close to $T_{SDW}$. A large increase of the forward scattering  between
carriers  that move in opposite  directions yields quite unfavorable conditions for  conventional BCS 
superconductivity
\cite{Bealmonod86}. These conditions become  even worst when the influence of interchain exchange $J_\perp$   is
included  in the analysis \cite{Bourbonnais88}.

\begin{figure}[htb]
\epsfxsize=0.9\hsize
\centerline{\epsfbox{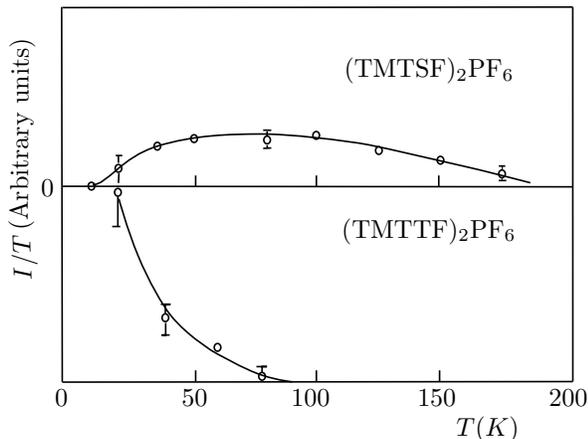}}
\caption{Temperature dependence of the $2k_F^0$ lattice susceptibility $(I/T)$ as function of temperature in
the normal phases of (TMTSF)$_2$PF$_6$ (top)  and (TMTTF)$_2$PF$_6$ (bottom), after \protect\cite{Pouget82}.}
\label{xray}
\end{figure}

 In the traditional BCS scheme  for superconductivity, the repulsion between
electrons  must be completely screened  by the attraction produced the  exchange  of  virtual phonons $-$ a
screening made possible only below the Debye temperature $\theta_D$ due to retardation. For narrow
tight-binding  bands   in the organics \cite{Barisic81}, the attraction is  strongest  for backscattering
processes in which $2{\bf k}_F$ phonons are exchanged.
According to the results of X-ray experiments performed on (TMTSF)$_2$X,
however,  the electron-phonon vertex part at this wavevector does not undergo any significant increase  in the
normal state \cite{Pouget82}. As it is indeed seen  in the upper-half of Figure~\ref{xray} for (TMTSF)$_2$PF$_6$,
the amplitude of the $2k_F^0$  lattice suceptibility $-$ which is directly involved in the strength of the phonon
exchange $-$ is weak (a result also found  for (TMTSF)$_2$ClO$_4$ \cite{Pouget82}). On this matter, it is
instructive to compare with the sulfur analog compound (TMTTF)$_2$PF$_6$ (Figure~\ref{xray}, bottom), for which
the electron-phonon vertex part at $2k_F^0$ becomes singular, signaling a lattice instability  towards a
spin-Peierls distortion (left-hand side of Fig.~\ref{Diagramme}). This instability produces a spin gap that is
clearly visible in the temperature dependence of  magnetic susceptibility  and nuclear relaxation rate
\cite{Creuzet87,Bourbon96}; these  effects are not seen in (TMTSF)$_2$X close to $P_c$. The persistent
enhancement of these quantities indicates that interactions are dominantly repulsive, making  the traditional
phonon-mediated source of pairing essentially inoperant in (TM)$_2$X. Added to that, it should be
stressed that the restoration of the SDW states under magnetic field above  $P_c$ gives an additional support of a
positive value for 
$g^*$ in this pressure range.

The above discussion naturally brings us to the question of the nature of pairing mechanism when electrons
 move in the presence of antiferromagnetic spin correlations. As suggested by Emery \cite{Emery86,Bealmonod86},
long-lived antiferromagnetic spin fluctuations of wavevector {\bf Q}$_0$ give rise to an oscillating potential in
real space at the corresponding periodicity,  to which electrons  are coupled. Electrons moving in opposite
directions can thus avoid local repulsion  and even be attracted  to each other if they move on different chains. 
The attraction being strongest for neighboring chains produces a gap with nodes on the Fermi surface.  This
mechanism can be seen as the spin-analog of
   the so-called Kohn-Luttinger mechanism proposed for  superconductivity induced by electrons   exchanging 
charge-density excitations $-$ as  produced by Friedel oscillations \cite{Kohn65}.

By virtue of the rather large temperature domain where spin correlations are present (Figure~\ref{Relaxation}),
one can expect that the K-L mechanism can emerge relatively deep in the normal phase  giving in turn rise to
interchain pairing correlations \cite{Bourbonnais88,Guay99}. Given the inability of the BCS nesting
mechanism to produce  spin correlations over a large temperature domain, it cannot alone 
account for pairing fluctuations far from $T_c$. But  interchain superexchange does. Following the analysis of
Ref.\cite{Guay99}, singlet interchain pairing correlations dominate  when intrachain SDW correlations couple
through the interchain superexchange. This is shown for instance by the following expression for the pairing
susceptibility in the normal state:
\begin{equation}
\chi^\perp_{i,j}(T) \sim [1-J_\perp(T)\chi(T)]^{-3/2},
\end{equation}
where $J_\perp(T)$ stands for an effective temperature dependent  transverse superexchange. It is worth noting
here that no (antisymmetric) interchain {\it triplet}  pairing correlations are found to be induced by SDW
fluctuations.  

As regards to the critical temperature $T_c$, some interesting features can be drawn in the framework of  the
spin-fluctuations  exchange mechanism. Adopting  the simple picture where a  Fermi
liquid component exists near
$P_c$ $-$ consistently  with the suppression of $T_{SDW}$ due to nesting deviations $-$, the `propagator' of spin
fluctuations involved in pairing attraction reads:
\begin{equation}
V({\bf Q},\omega) = {V_0 \over  r + \sum_ i\xi^2_{0,i} (Q_i- Q_0)^2 -i\omega/\omega_0},
\end{equation}
where $ r= [T-T_{SDW}(P)]/T_{SDW}(P)$,  $\xi_{0,i}$ is the SDW coherence length in the $i=x,y,z$ directions, and
$\omega_0 \sim T_{SDW}$ is a characteristic frequency scale for spin fluctuations.
The critical temperature for superconductivity can
be calculated  within a standard BCS procedure. The
result       for $T_c$ can be written in the familiar form:
\begin{equation}
T_c \sim \omega_0\  e^{-1/\lambda^*},
\end{equation}
where in the usal way, $\omega_0$ enters as a cut-off for $T_c$ due to retardation. The effective coupling
constant
$\lambda^* \propto 1/\kappa $ involves a `stiffness' constant $\kappa$ ($\kappa \to 0$ as  $r \to 0$) of the SDW
fluctuations. Therefore as one approaches $P_c$, $r$ becomes small so that $\lambda^*$ reaches strong coupling 
but $\omega_0$ remains  small. In these conditions, the growth of $T_c$ levels off  and  reaches a maximum in
agreement with the maximum commonly seen at $P_c$ (Figure~\ref{Diagramme}).  This can be seen as a limitation of
$T_c$ in (TM)$_2$X, a situation that, bears an obvious ressemblance with the limitation of $T_c$ in 
conventional superconductors,  where  the increase of $T_c$ reaches a maximum  when the elasticity of
the lattice becomes too soft and the coupling becomes strong. In the present spin-fluctuation scheme, the 
 maximum $T_c$ would therefore scale with
$\omega_0$.  Above
$P_c$, the amplitude of fluctuations decreases (Figure~\ref{Relaxation}), so that
$T_c$ is dominated by a reduction of $\lambda^*$ and therefore becomes a decreasing function of
pressure \cite{Caron86}.

In conclusion, the existence of spin correlations in  (TM)$_2$X near the critical pressure
for superconductivity  indicates that the normal phase precursor to this broken-symmetry  state is strongly
correlated and would make  the conventional electron-phonon mechanism inoperant for the origin of this phase.
Interchain superexchange is set to play an important role in these systems not only in the stabilization of
long-range magnetic ordering but also as a mechanism  that magnifies pairing induced by spin-fluctuation
exchange. 

\noindent
{\bf Acknowledgements} The author thanks D.~J\'erome, L. G. Caron, B. Guay and R. Duprat for
valuable discussions on several aspects of this brief review. The author would also like to thank  D.
S\'en\'echal for very useful comments. This work is supported by NSERC of
Canada and the Superconductivity program of the Canadian Institute of  Advanced Research (CIAR).

\bibliography{articles}

\begin{thebibliography}{10}

\bibitem{Jerome80}
D. J\'erome, A. Mazaud, M. Ribault, and K. Bechgaard, J. Phys. (Paris) Lett.
  {\bf 41},  L95  (1980).

\bibitem{Jerome82}
D. J\'erome and H. Schulz, Adv. in Physics {\bf 31},  299  (1982).

\bibitem{Kanoda97}
K. Kanoda, Physica C {\bf 282-287},  299  (1997).

\bibitem{McKenzie98}
R.~H. McKenzie, Comments Cond. Matt. Phys. {\bf 18},  309  (1998).

\bibitem{Lefebvre00}
S. Lefebvre {\it et~al.}, this volume and to be published.

\bibitem{Bourbon99}
C. Bourbonnais and D. J\'erome,  in {\em Advances in Synthetic Metals, Twenty
  Years of Progress in Science and Technology}, edited by P. Bernier, S.
  Lefrant, and G. Bidan (Elsevier, New York, 1999), pp.\ 206--261.

\bibitem{Yamaji82}
K. Yamaji, J. Phys. Soc. of Japan {\bf 51},  2787  (1982).

\bibitem{Gorkov84}
L.~P. Gorkov and A.~G. Lebed, J. Phys. (Paris) Lett. {\bf 45},  L433  (1984).

\bibitem{Heritier84}
M. H\'eritier, G. Montambaux, and P. Lederer, J. Phys. (Paris) Lett. {\bf 45},
  L943  (1984).

\bibitem{Jerome94}
D. J\'erome,  in {\em Organic Conductors: fundamentals and applications},
  edited by J.-P. Farges (Dekker, New York, 1994), pp.\ 405--494.

\bibitem{Chaikin96}
P. Chaikin, J. Phys. I (France) {\bf 6},  1875  (1996).

\bibitem{Creuzet87b}
F. Creuzet {\it et~al.}, Synthetic Metals {\bf 19},  277  (1987).

\bibitem{Bourbon84}
C. Bourbonnais {\it et~al.}, J. Phys. (Paris) Lett. {\bf 45},  L755  (1984).

\bibitem{Georges00}
A. Georges, T. Giamarchi, and N. Sandler, preprint cond-mat/0001063
  (unpublished).

\bibitem{Danner94}
G.~M. Danner, W. Kang, and P.~M. Chaikin, Phys. Rev. Lett. {\bf 72},  3714
  (1994).

\bibitem{Moser98}
J. Moser {\it et~al.}, Eur. Phys. J. B {\bf 1},  39  (1998).

\bibitem{Zheleznyak99}
A.~T. Zheleznyak and V.~M. Yakovenko, Eur. Phys. J. B {\bf 11},  385  (1999).

\bibitem{Coulon82}
C. Coulon {\it et~al.}, J. Phys. (Paris) {\bf 43},  1059  (1982).

\bibitem{Emery82}
V.~J. Emery, R. Bruisma, and S. Barisic, Phys. Rev. Lett. {\bf 48},  1039
  (1982).

\bibitem{Canadell94}
L. Balicas {\it et~al.}, J. Phys. I (France) {\bf 4},  1539  (1994).

\bibitem{Creuzet87}
F. Creuzet {\it et~al.}, Synthetic Metals {\bf 19},  289  (1987).

\bibitem{Wzietek93}
P. Wzietek {\it et~al.}, J. Phys. I (France) {\bf 3},  171  (1993).

\bibitem{Bourbon93}
C. Bourbonnais, J. Phys. I (France) {\bf 3},  143  (1993).

\bibitem{Emery79}
V.~J. Emery,  in {\em Highly Conducting One-Dimensional Solids}, edited by
  J.~T. Devreese, R.~E. Evrard, and V.~E. van Doren (Plenum Press, New York,
  1979), p.\ 247.

\bibitem{Brazovskii85b}
S. Brazovskii and Y. Yakovenko, Sov. Phys. JETP {\bf 62},  1340  (1985).

\bibitem{Bourbon86}
C. Bourbonnais and L.~G. Caron, Physica {\bf 143B},  450  (1986).

\bibitem{Bourbonnais88}
C. Bourbonnais and L. Caron, Europhys. Lett. {\bf 5},  209  (1988).

\bibitem{Bourbon95}
C. Bourbonnais,  in {\em Les Houches, Session LVI (1991), Strongly interacting
  fermions and high-T$_c$ superconductivity}, edited by B. Doucot and J.
  Zinn-Justin (Elsevier Science, Amsterdam, 1995), p.\ 307.

\bibitem{Klemme95}
B.~J. Klemme {\it et~al.}, Phys. Rev. Lett. {\bf 75},  2408  (1995).

\bibitem{Brown97}
S.~E. Brown {\it et~al.}, Synthetic Metals {\bf 86},  1937  (1997).

\bibitem{Vescoli98}
V. Vescoli {\it et~al.}, Science {\bf 281},  1181  (1998).

\bibitem{Timusk96}
N. Cao, T. Timusk, and K. Bechgaard, J. Phys. I (France) {\bf 6},  1719
  (1996).

\bibitem{Giamarchi97}
T. Giamarchi, Physica {\bf B230-232},  975  (1997).

\bibitem{Bealmonod86}
M.~T. B\'eal-Monod, C. Bourbonnais, and V.~J. Emery, Phys. Rev. B {\bf 34},
  7716  (1986).

\bibitem{Pouget82}
J. Pouget {\it et~al.}, Mol. Cryst. Liq. Cryst. {\bf 79},  129  (1982).

\bibitem{Barisic81}
S. Barisic and S. Brazovskii,  in {\em Recent Developments in Condensed Matter
  Physics}, edited by J.~T. Devreese (Plenum, New York, 1981), Vol.~1, p.\ 327.

\bibitem{Bourbon96}
C. Bourbonnais and B. Dumoulin, J. Phys. I (France) {\bf 6},  1727  (1996).

\bibitem{Emery86}
V.~J. Emery, Synthetic Metals {\bf 13},  21  (1986).

\bibitem{Kohn65}
W. Kohn and J.~M. Luttinger, Phys. Rev. Lett. {\bf 15},  524  (1965).

\bibitem{Guay99}
B. Guay and C. Bourbonnais, Synthetic Metals {\bf 103},  2180  (1999).

\bibitem{Caron86}
L.~G. Caron and C. Bourbonnais, Physica {\bf 143B},  453  (1986).

\end{thebibliography}
\bibliographystyle{prsty}
\end{document}